\documentclass[11pt,a4paper]{article}
\usepackage{jheppub}
\usepackage{tabularx,afterpage,xifthen}
\usepackage{amsmath,amsfonts,amssymb,mathtools}
\usepackage{slashed}
\usepackage{comment}
\usepackage{tikz}
\usepackage{graphicx}
\usepackage{lmodern}
\usepackage{xcolor}
\usepackage[compat=1.1.0]{tikz-feynman}
\usepackage{xsavebox}
\usepackage{subcaption}

\makeatletter
\gdef\@fpheader{\phantom{Prepared for submission to JHEP}}
\makeatother
\newcommand{\Lambdah}{\Lambda_{\rm h}}

\title{A new effective field theory for heavy quarks in the quark-gluon plasma}
\author[]{Andreas Kirchner,}
\emailAdd{andreas.kirchner@duke.edu}
\author[]{Berndt M{\"u}ller,}
\emailAdd{muller@phy.duke.edu}
\author[]{Jyotirmoy Roy,}
\emailAdd{jyotirmoy.roy@duke.edu}
\author[]{Chathuranga Sirimanna}
\emailAdd{chathuranga.sirimanna@duke.edu}
\affiliation[]{Department of Physics, Duke University, Durham, NC 27708, USA }

\abstract{An effective field theory framework is developed to study the interaction of heavy quarks in strongly coupled quark-gluon plasma (QGP). The latter is treated as a relativistic non-dissipative colorless fluid which can be studied using a derivatively coupled effective field theory based on previous work. Coupling this to heavy quarks provides a systematic way to obtain the interaction between the heavy quark and phonons, excitations of the fluid. In particular we calculate the quasiparticle width of heavy quark to phonon and phonon-heavy quark scattering in a thermal medium.}

\begin{document}

\maketitle

\section{Introduction}
\label{sec:Intro}

The central paradigm of relativistic heavy ion physics is that the quark-gluon plasma (QGP) behaves as a nearly ``perfect'' fluid, i.~e.\ as a fluid with kinematic viscosity near the quantum limit ($\eta_{\rm kin} = \eta/s \approx 1/4\pi$.) Hydrodynamics, the theory governing fluid motion, is applicable to all energy-conserving, interacting many-body systems on long distance scales. By virtue of focusing on the description of the behavior of such a system at long-distance, low-momentum scales, hydrodynamics fits well into the general concept of effective field theories (EFTs) that capture the dynamical properties of a many-body system within a certain domain of scales without the need to explicitly treat the system microscopically. The microscopic dynamics are only manifest in the coefficients of the relevant terms in the effective action that defines the EFT. 

The essential aspects governing the construction of an EFT are the symmetries obeyed by the effective action, the domain of validity of the EFT expressed by a characteristic upper momentum scale $\Lambdah$, and a power counting scheme that lets us order allowed terms in the action by their relative magnitude. Such an EFT framework has recently been developed for relativistic hydrodynamics \cite{Endlich:2010hf,Dubovsky:2011sj,Grozdanov:2013dba,Montenegro:2016gjq,Liu:2018kfw, Haehl:2018lcu}. Here we aim to go one step further and extend this to a quark-hydro effective field theory (QHEFT) by considering hard probes that are not themselves part of the fluid but interact with it. There has been some work done in regards to this \cite{Gupta:2023tue} but a complete picture with all the operators taken into account is still missing. To be amenable to such a description, the probes have to be either weakly interacting so that they do not equilibrate on the relevant time scales (electromagnetic and weakly interacting probes come to mind) or must be characterized by a hard scale, larger than the scale of the collective modes. Examples of the latter type of probes of a quark-gluon plasma are heavy quarks with mass $M \gg T$ and energetic quarks or gluons with $E \gg \Lambdah$ that give rise to jets. Because of their importance in relativistic heavy ion collision phenomenology and the relative ease with which they can be coupled to the QGP fluid we begin this exploration with a treatment of heavy quarks.

Because of their large mass, multiple interactions with QGP constituents are required to appreciably change the momenta of heavy quarks. We can therefore think of this interaction as being between the QGP fluid and the heavy quark rather than between the quark and an individual QGP constituent. This picture applies as long as the momentum exchange between the QGP and the heavy quark lies within the domain of validity of the QHEFT description, i.~e.\ when $|\vec{q}| < \Lambdah$, where $\vec{q}$ is the momentum transfer. Whether this collective interaction with the fluid is the dominant influence on the motion of the heavy quark depends on the ratio $|\vec{q}|/\Lambdah$. If $|\vec{q}| \gg \Lambdah$, interactions of the heavy quark with the microscopic constituents of the medium dominate over the collective quark-fluid interactions. Conversely, when $\Lambdah \ge |\vec{q}|$, collective quark-fluid interactions are the dominant mechanism for the heavy quark to interact with the fluid.

The upper momentum scale $\Lambdah$ can be related to $T$, the temperature of the QGP. The ratio $\Lambdah/T$ is determined by the strength of the coupling among the microscopic constituents of the fluid. Within the framework of kinetic theory, $\Lambdah \propto \lambda_{\rm tr}^{-1}$, where $\lambda_{\rm tr} = (\rho\int dq^2 (d\sigma/dq^2)\sin^2\theta_q)^{-1}$ is the inverse of the transport mean free path $\lambda_{\rm tr}$ that is defined in terms of the differential cross section $d\sigma/dq^2$, the scattering angle $\theta_q$, and the fluid density $\rho$. In weakly coupled gauge theories $\lambda_{\rm tr}^{-1} \sim g^4\ln(1/g)T \ll T$ \cite{Arnold:2002zm}, which means that the fluid plays a subordinate role in the description of heavy quark motion. At strong coupling the value of $\Lambdah$ is constrained by the first nonhydrodynamic quasinormal mode \cite{Kovtun:2005ev,Berti:2009kk,DeLescluze:2025jqx}, which usually is a modest multiple of $\pi T$, and the interaction with the fluid can be the dominant mechanism for momentum exchange between the quark and the surrounding medium.

Even when $|\vec{q}| < T$, the momentum transfer between the quark and the QGP fluid can be phenomenologically relevant, if one can observe the response of the fluid to the passage of the heavy quark. This also applies to the response of the QGP fluid to the passage of a highly energetic light quark or gluon and its evolving parton shower. Such phenomena have been the subject of theoretical speculation for some time \cite{Stoecker:2004qu,Casalderrey-Solana:2004fdk,Ruppert:2005uz,Renk:2005si,Chaudhuri:2005vc,Betz:2010qh,Bouras:2014rea,Casalderrey-Solana:2020rsj,Yang:2021qtl,Yang:2022nei}, but they remained unobserved until recently \cite{CMS:2025dua}. Microscopic calculations of the structure of the energy-momentum deposition by an energetic quark into the QGP have been performed for strongly coupled gauge theories using holographic techniques \cite{Friess:2006fk,Gubser:2007ga,Chesler:2007sv} and in QCD using thermal perturbation theory \cite{Neufeld:2008hs,Neufeld:2010xi}. These calculations can be used to derive microscopic predictions for the Wilson coefficient of the quark-fluid interaction term introduced here.

This manuscript is structured as follows: In Section \ref{sec:HEFT} we briefly review the effective field theory formulation of relativistic hydrodynamics and discuss the hydrodynamic scale setting. We also discuss the emergence of phonons as propagating degree of freedom of the QHEFT and the coupling of quarks and gluons to the fluid. In Section \ref{sec:QPhscattering} we evaluate the interaction between quarks and phonons in a thermal fluid, calculate the thermal quark decay rate, and derive the quark-phonon scattering cross section. The manuscript concludes with a brief summary and an outlook on ongoing and future work.

\section{QHEFT Framework}
\label{sec:HEFT}

\subsection{Construction of the Hydrodynamic EFT}
\label{sec:HEFT_construction}

The effective field theory for non-dissipative fluids has been written down in several works, which we will review here \cite{Endlich:2010hf,Dubovsky:2011sj,Grozdanov:2013dba}. The low energy hydrodynamical degrees of freedom for this system are taken to be the comoving coordinates of fluid volume elements as a function of their physical position $\vec{x}$ at time $t$. These can be represented by 3-dimensional scalar fields 
\begin{equation}
\phi_I=\phi_I(t,\vec{x}), \; I=1,2,3 .   
\end{equation}
With the given degrees of freedom, one should be able to construct all the operators of the theory based on symmetries and power counting of the theory. The former can be ascertained from the ground state of the system, which in this case is defined to be the configuration of a stationary fluid in its own rest frame, for which 
\begin{equation}
    \langle \phi_{I}(t,\vec{x}) \rangle = x_I
\end{equation}
for all times. While the ground state can spontaneously break some of the symmetries of the Lagrangian, the residual symmetries of the ground state necessarily are the symmetries of the Lagrangian. Hence, one can use these to identify the internal symmetries of the Lagrangian, which in this case corresponds to
\begin{align}
    & \phi_{I} \rightarrow \phi_{I}+a_{I}, \quad a_{I}=\text{const}, \label{Eq:TransInv}\\
    & \phi_{I} \rightarrow R_{I}^{J} \phi_{J}, \quad R \in SO(3), \label{Eq:RotInv}\\
    & \phi_{I} \rightarrow \xi_{I} (\phi), \quad \text{det}(\partial \xi_{I} /\partial \phi_{J})=1. \label{Eq:PCInv}
\end{align}
Eqns.~(\ref{Eq:TransInv}) and (\ref{Eq:RotInv}) correspond to the ground state being translation and rotation invariant. The symmetry in Eqn.~(\ref{Eq:PCInv}) distinguishes a perfect fluid from an isotropic solid (jelly) by the absence of a shape restoring force, whereby the energy of the system is unchanged under arbitrary volume-preserving deformations. The number of derivatives acting on the scalar fields systematically accounts for the gradient expansion of hydrodynamics. The minimal term in this gradient expansion that respects the symmetries is the vector field
\begin{align}
    \label{Eq:Current}
    J^{\mu} = \frac{1}{3!}\varepsilon^{\mu \nu \rho \sigma} \varepsilon_{I J K} \partial_{\nu} \phi^{I} \partial_{\rho} \phi^{J} \partial_{\sigma} \phi^{K}.
\end{align}
To write down the Lagrangian, we need to construct Lorentz scalars that are lowest order in the derivative expansion. The simplest Lorentz scalar we need to consider is 
\begin{equation}
    X=J^{\mu}J_{\mu}.
\end{equation}
The lowest order Lagrangian for the fluid can then be written as:
\begin{equation}
\label{Eq:PhononLagrangian}
   \mathcal{L} =F(X).
\end{equation}
where $F$ is an arbitrary function of the dimensionless Lorentz scalar $X$. This corresponds to the energy-momentum tensor of perfect fluid
\begin{eqnarray}
    T^{\mu \nu}=(p+\varepsilon)u^\mu u^\nu -p \,\eta^{\mu \nu},
\end{eqnarray}
with the identification of the fluid energy density and pressure to be
\begin{align}
 \varepsilon=-F(X),  \qquad   p=F(X)-2X F'(X),
\end{align}
where $u^\mu=J^{\mu}/\sqrt{X}$ defines the fluid velocity.
The equation of motion for the fluid is given by the conservation law
\begin{equation}
   \partial_\sigma \left[F'(X) \frac{1}{2} \varepsilon^{\sigma \alpha \beta \gamma} \varepsilon_{IJK} J_{\alpha} \partial_{\beta}\phi^{J} \partial_{\gamma}\phi^{K} \right]=0
\end{equation}

In this section, ideal fluid dynamics has been formulated as a classical field theory. As classical ideal fluid dynamics does not have an intrinsic scale, all terms contained in the function $F(X)$ are of same order in the derivative expansion. As a next step we will introduce fluctuations of the fluid field around its equilibrium value, which will be quantized. As in any field theory with quantized modes, we will use the canonical normalization of their contribution to the action. This normalization naturally introduces a suppression scale (the ``hydro scale'' $\Lambda_h$) for higher derivative terms in the expansion of the functions appearing in the Lagrangian. This analysis elevates all QCD couplings to functions of the dimensionless variable $X$ with Wilson coefficients that are suppressed by appropriate powers of the new scale.

\subsection{Fluid excitations: Phonons}
\label{sec:Phonons}

\begin{figure}[!ht]
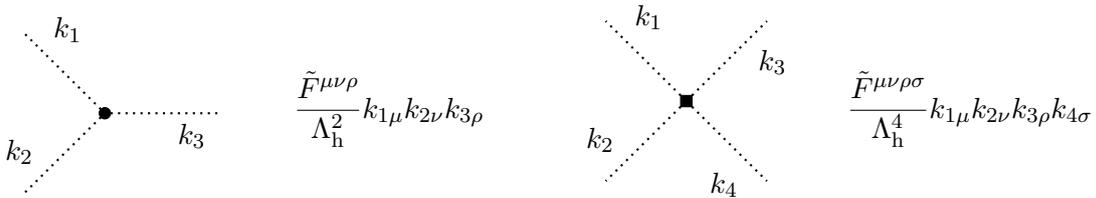

    \begin{subfigure}{0.5\textwidth}
        \begin{minipage}{0.35\textwidth}
            \centering
            \xusebox{3Phonon}
        \end{minipage}
        \hspace{1em}
        \begin{minipage}[t]{0.5\textwidth}
            \vspace{-0.9cm} 
            \[
                \frac{\Tilde{F}^{\mu \nu\rho}}{\Lambdah^2} k_{1\mu} k_{2\nu} k_{3\rho}
            \]
        \end{minipage}
    \end{subfigure}
    \begin{subfigure}{0.5\textwidth}
        \begin{minipage}{0.35\textwidth}
            \centering
            \xusebox{4Phonon}
        \end{minipage}
        \hspace{1em}
        \begin{minipage}[t]{0.5\textwidth}
            \vspace{-0.9cm} 
            \[
                \frac{\Tilde{F}^{\mu \nu\rho \sigma}}{\Lambdah^4} k_{1\mu} k_{2\nu} k_{3\rho} k_{4\sigma}
            \]
        \end{minipage}
    \end{subfigure}
    \caption{Feynman diagrams for three and four phonon self-interactions. $n$-phonon interactions exist for all $n>2$, but are suppressed by $\Lambdah^{-(2n-4)}$, respectively.}
    \label{fig:FeynmanRulesPhonons}
\end{figure}

We will now discuss the displacement fluctuation spectrum of the theory and show that it is described by phonon degrees of freedom. As a first step we allow the fluid field to vary around its equilibrium value,
\begin{equation}
    \phi^I = x^I + \lambda \pi^I(t,\vec{x}),
\end{equation}
with $ \lambda $ being a dimensionless expansion parameter which will be set to $\lambda \to 1$ at the end of the expansion and $ \pi^I(t,\vec{x}) $ being the displacement from equilibrium. Applying this expansion to the 
current (\ref{Eq:Current}), we find
\begin{align}
    J^\mu 
    &= J^\mu_{(0)} + \lambda J^\mu_{(1)} + \lambda^2 J^\mu_{(2)} + \lambda^3 J^\mu_{(3)}
\end{align}
with\footnote{A Levi-Civita tensor with mixed Greek and Roman letters is to be understood as $\epsilon_{\mu \nu \rho I}=\epsilon_{\mu \nu \rho \sigma} \delta_I^\sigma$.}
\begin{align}
    J^\mu_{(0)} &= b^\mu = (1,0,0,0) \quad (\text{rest frame}), \label{Eq:J0}\\
    J^\mu_{(1)} &= \frac{1}{2} \epsilon^{\mu \nu\rho\sigma} \epsilon_{\nu\rho K} \partial_\sigma \pi^K, \label{Eq:J1}\\
    J^\mu_{(2)} &= \frac{1}{2} \epsilon^{\mu \nu\rho\sigma} \epsilon_{\nu JK} \partial_\rho \pi^J \partial_\sigma \pi^K, \\
    J^\mu_{(3)} &= \frac{1}{6} \epsilon^{\mu\nu\rho\sigma} \epsilon_{IJK} \partial_\nu  \pi^I \partial_\rho \pi^J \partial_\sigma \pi^K.
 \end{align}
Here, the vector $b^\mu$ encodes the (frame dependent) four-velocity of the fluid background. The last term in Eq.~(\ref{Eq:J0}) reflects our choice to work in the fluid rest frame.\footnote{Note that $b^\mu = (1,0,0,0)$ in the fluid rest frame is a consequence of our choice of normalization for $J^\mu$.} Note that each $J^\mu_{(n)}$ includes $n$ displacement fields $\pi^I$. Applying the expansion of $J^\mu$ to the scalar $X=J^\mu J_\mu$, we find
\begin{align}
    X &= J^\mu_{(0)} J_{(0)\mu} + \lambda [ 2 J^\mu_{(0)} J_{(1)\mu}] +\lambda^2[2J^\mu_{(0)} J_{(2)\mu} +J^\mu_{(1)} J_{(1)\mu}] +\lambda^3[2J^\mu_{(0)} J_{(3)\mu} +2J^\mu_{(2)} J_{(1)\mu}]
    \nonumber\\
    &+ \lambda^4[2J^\mu_{(2)} J_{(2)\mu} + 2J^\mu_{(1)} J_{(3)\mu}] + \lambda^5 2 J^\mu_{(2)} J_{(3)\mu} + \lambda^6 J^\mu_{(3)} J_{(3)\mu}
\end{align}
The series shown here is finite, due to the truncation of $J^\mu$ to the lowest order operator in derivative expansion, but in general would have higher order terms. The expansion of $F(X)$ around the background is then given by 
\begin{align}
    F(X)&= F(1) + 2 b_\mu J_{(1)}^\mu F'(1) \lambda \nonumber\\
    &+ \Big[(J^\mu_{(1)} J_{(1)\mu}+ 2 b_\mu J^\mu_{(2)})F'(1) + 2 (b_\mu J^\mu_{(1)} )^2F''(1) \Big] \lambda^2 \nonumber\\
    &+ \Big[2(J^\mu_{(1)}J_{(2)\mu} + b_\mu J^\mu_{(3)})F'(1) + 2b_\mu J^\mu_{(1)}(J^\mu_{(1)}J_{(1)\mu}+2b_\mu J^\mu_{(2)})F''(1) \nonumber\\
    &+ \frac{4}{3} (b_\mu J^\mu_{(1)})^3  F'''(1) \Big] \lambda^3 + \mathcal{O}(\lambda^4)
\label{Eq:ExpansionF}
\end{align}
Employing the form of the current Eq.~(\ref{Eq:J1}), we can neglect the term linear in $\lambda$, since $b_\mu J_{(1)}^\mu=b^{0} (\Vec{\nabla} \cdot \Vec{\pi}) - \Vec{b} \cdot \partial_{t} \Vec{\pi}$ is a total derivative. Evaluating the remaining terms explicitly, dropping the constant terms and taking $\lambda \to 1$, we find the final form of the Lagrangian up to third order in derivatives:
\begin{align}
    \mathcal{L}= F^{\mu \nu}_{IJ} \partial_\mu \pi^I \partial_\nu \pi^J + F^{\mu \nu\rho}_{IJK} \partial_\mu \pi^I \partial_\nu \pi^J \partial_\rho \pi^K + \mathcal{O}((\partial \pi)^4)
\end{align}
with the coefficients
\begin{eqnarray}
    F^{\mu \nu}_{IJ} &=& \left[ \frac{1}{4} \epsilon^{\alpha_1 \alpha_2 \alpha_3 \mu} \epsilon_{\alpha_1}^{\alpha_2' \alpha_3'\nu} \epsilon_{\alpha_2 \alpha_3 I} \epsilon_{\alpha_2' \alpha_3' J} + \epsilon^{0 \alpha \mu \nu} \epsilon_{\alpha IJ} \right] F'(1) \nonumber\\
    &&+ \frac{1}{2} \epsilon^{0\alpha_1\alpha_2\mu} \epsilon^{0\alpha_1'\alpha_2'\nu} \epsilon_{\alpha_1 \alpha_2 I} \epsilon_{\alpha_1' \alpha_2' J}F''(1), \\
    F^{\mu \nu \rho}_{IJK} &=& \left[\frac{1}{2} \epsilon^{\alpha_1 \alpha_2 \alpha_3 \mu } \epsilon_{\alpha_1}^{\alpha_2' \nu\rho} \epsilon_{\alpha_1\alpha_2 I } \epsilon_{\alpha_2' JK} + \frac{1}{3}\epsilon^{0\mu\nu\rho}\epsilon_{IJK}\right]F'(1) \nonumber\\
    &&+\frac{1}{2} \epsilon^{0\alpha_1\alpha_2\mu}  \epsilon_{\alpha_1 \alpha_2 I}  [\frac{1}{4} \epsilon^{\alpha_3 \alpha_4 \alpha_5 \nu} \epsilon_{\alpha_3}^{\alpha_4' \alpha_5'\rho} \epsilon_{\alpha_4 \alpha_5 J} \epsilon_{\alpha_4' \alpha_5' K} + \epsilon^{0 \alpha \nu \rho} \epsilon_{\alpha JK}]F''(1) \nonumber\\
    &&+ \frac{1}{6} \epsilon^{0\alpha_1\alpha_2\mu} \epsilon^{0\alpha_1'\alpha_2'\nu} \epsilon^{0\alpha_1''\alpha_2''\rho} \epsilon_{\alpha_1 \alpha_2 I} \epsilon_{\alpha_1' \alpha_2' J}
    \epsilon_{\alpha_1'' \alpha_2'' K} F'''(1).
\end{eqnarray}
Here we want to highlight two points: Due to the definition of the current $J^\mu$, every term of the Lagrangian can be expressed as $F^{\mu_1 \ldots \mu_n}_{I_1 \ldots I_n} \partial_{\mu_1} \pi^{I_1} \ldots \partial_{\mu_n} \pi^{I_n}$, with 
$F^{\mu_1 \ldots \mu_n}_{I_1 \ldots I_n}$ being a complicated expression containing $F(1)$ and its derivatives up to $n$-th order as well as various combinations of Levi-Civita symbols. Collecting all the quadratic terms gives
\begin{align}
\label{Eq:GeneralPhononLagrangian}
    \mathcal{L}^{(2)} &= -F'(1) \left[ \frac{1}{2}(\partial_t \vec{\pi})^2 - \frac{1}{2}\frac{F'(1)+2F''(1)}{F'(1)} (\nabla \cdot \vec{\pi})^2 \right],
 \end{align} 
which can be identified with the kinetic term for the displacement fields that propagate with the speed of sound $c_s^2=\frac{F'(1)+2F''(1)}{F'(1)}$. Note that $\mathcal{L}^{(2)}$ is the product of the expansion of the generic function $F(X)$ and therefore is a result that will be used again when coupling with other degrees of freedom. This kinetic term is not canonically normalized and therefore a field redefinition given by
\begin{equation}
\label{Eq:PhononRescaling}
    \pi^I \to \sqrt{-F'(1)}\pi^I
\end{equation}
results in the following form of the Lagrangian density
\begin{align}
    \mathcal{L}^{(2)} &= \frac{1}{2}(\partial_t \vec{\pi})^2 -\frac{1}{2}c_s^2(\nabla \cdot \vec{\pi})^2.
 \end{align}
There are several points to be discussed here. Firstly this quadratic Lagrangian corresponds to three independent displacement fields which can be separated into transverse and longitudinal components $\vec{\pi}=\vec{\pi}_{L}+\vec{\pi}_{T}$, determined by the conditions $\nabla \times \vec{\pi}_{L}=0$ and $\nabla \cdot \vec{\pi}_{T}=0$. The longitudinal components admit wave solutions with the dispersion relation $E=c_s |\vec{p}|$ and correspond to phonons. The transverse components represent shear modes with the dispersion relation $E=0$ since the described fluid is ideal. For a dissipative fluid, this dispersion relation will change. Secondly, with the field redefinition, the displacement field now has mass dimension $+1$ instead of $-1$. Applying this redefinition to higher-order terms
introduces a natural scale, by which higher-order terms are suppressed
\begin{equation}
\label{Eq:Dim6Operator}
    F^{\mu \nu\rho}_{IJK} \partial_\mu \pi^I \partial_\nu \pi^J \partial_\rho \pi^K \to \frac{F^{\mu \nu\rho}_{IJK}}{(-F'(1))^{3/2}} \partial_\mu \pi^I \partial_\nu \pi^J \partial_\rho \pi^K.
\end{equation}
To write it in the form of higher-dimensional operators in EFT, we introduce dimensionless Wilson coefficients $\Tilde{F}^{\mu\nu\rho}_{IJK}$\footnote{Here and in the following, the tilde version of a quantity is always the dimensionless version of the original quantity, normalized by $\Lambdah$.}, defined as
\begin{align}
    \Tilde{F}^{\mu\nu\rho}_{IJK} &= F^{\mu\nu\rho}_{IJK}/(-F'(1)).
\end{align}
Thus Eqn.~(\ref{Eq:Dim6Operator}) reduces to
\begin{equation}
 \frac{\Tilde{F}^{\mu \nu\rho}_{IJK}}{\sqrt{-F'(1)}} \partial_\mu \pi^I \partial_\nu \pi^J \partial_\rho \pi^K.
\end{equation}
Being a dimension 6 operator, this term can then be used to identify the UV scale to be
\begin{equation}
    \Lambda_h =\sqrt[4]{-F'(1)}.
\end{equation}
The higher-order phonon self-interactions can then be derived similarly, with the Feynman rules for the 3 and 4 phonon self-interactions being shown in Fig.~\ref{fig:FeynmanRulesPhonons} where the 4 phonon interaction is suppressed by $\Lambdah^2$ compared to the 3 phonon interaction. The determination of this UV scale is discussed in the next section. 
\begin{xlrbox}{3Phonon}
  \begin{tikzpicture}
    \begin{feynman}
      \vertex (a);
      \vertex [dot][right=of a] (b){};
      \vertex [above left=of b] (g1);
      \vertex [below left=of b] (g2);
      \vertex [right=1.5cm of b] (g3);

      \diagram*{
        (b) -- [ghost, edge label'=\(k_1\), near end] (g1),
        (b) -- [ghost, edge label'=\(k_2\), near end] (g2),
        (b) -- [ghost, edge label'=\(k_3\), near end] (g3),
      };
    \end{feynman}
  \end{tikzpicture}
\end{xlrbox}

\begin{xlrbox}{4Phonon}
  \begin{tikzpicture}
    \begin{feynman}
      \vertex (a);
      \vertex [square dot][right=of a] (b){};
      \vertex [above left=of b] (g1);
      \vertex [below left=of b] (g2);
      \vertex [above right=of b] (g3);
      \vertex [below right=of b] (g4);

      \diagram*{
        (b) -- [ghost, edge label'=\(k_1\), near end] (g1),
        (b) -- [ghost, edge label'=\(k_2\), near end] (g2),
        (b) -- [ghost, edge label'=\(k_3\), near end] (g3),
        (b) -- [ghost, edge label'=\(k_4\), near end] (g4),
      };
    \end{feynman}
  \end{tikzpicture}
\end{xlrbox}

\subsection{Hydrodynamic Scale}
\label{sec:Scale}

One important question is the range of scales over which the hydrodynamic EFT applies. This can be addressed in multiple ways. One is within the EFT framework itself. For a bottom-up EFT the UV scale cannot be exactly determined since the UV theory is either not known or it is not possible to analytically match the low energy theory to the UV theory. But one can still ascertain the UV scale from symmetry and power counting arguments. The operators allowed in the theory are constrained by the symmetry, while the power counting dictates the order of importance of the different operators. Power counting dictates that the non-renormalizable operators are suppressed by the UV scale of the theory. Thus, identifying these operators helps in extracting this scale. As we have shown in Section \ref{sec:Phonons}, using this methodology for the hydrodynamic EFT, we find the UV scale to be related to the energy density of the fluid. 

The other approach is by considering the momentum domain in which the microscopic theory exhibits hydrodynamic behavior. There are two criteria that can be applied: (1) The momentum range over which the phonon mode behaves as a quasiparticle mode. (2) The energy scale at which non-hydrodynamical modes begin to control the relaxation of perturbations to equilibrium. We now consider these two criteria at weak coupling (perturbative thermal QCD) and for strong coupling (${\cal N}=4$ super-Yang-Mills theory for 't Hooft coupling $\lambda \gg 1$).

The dynamical limit of phonon propagation is given by the sound attenuation length
\begin{equation}
    k \ll \ell_s^{-1} = \frac{\left(\frac{4}{3}\eta+\zeta\right)k^2}{2c_s (\varepsilon+p)} ,
\end{equation}
where $c_s$ is the speed of sound, $\varepsilon+p = sT$ with $s$ being the entropy density, and $\eta,\zeta$ are the shear and bulk viscosity, respectively. Neglecting the bulk viscosity, we obtain the condition
\begin{equation}
    k \ll \frac{3c_s}{2\eta/s}T .
\end{equation}
At next-to-leading order, the kinematic shear viscosity for the thermal coupling $g(T) \approx 3.2$ is $\eta/s \approx 0.14$ \cite{Ghiglieri:2018dib}. With the estimate $c_s \approx 0.5$ \cite{Andersen:2014dua} we obtain the constraint $k \lesssim 5\, T$.

The limitation on the phonon mode in the strongly coupled ${\cal N}=4$ super-Yang-Mills theory is the same as in QCD, except that the kinematic shear viscosity at strong coupling at next-to-leading order is given by \cite{Buchel:2008sh}
\begin{equation}
    \eta/s = \frac{1}{4\pi}(1+15\zeta(3)\lambda^{-3/2}) 
\end{equation}
and the speed of sound is $c_s = 1/\sqrt{3}$. This results in the constraint 
\begin{equation}
    k \ll 2\sqrt{3}\pi T \approx 11\, T ,
\end{equation}
at infinite 't Hooft coupling. Alternatively, we can constrain the range of phonon momenta by the first non-hydrodynamical quasinormal mode in the shear channel \cite{Waeber:2015oka}: 
\begin{equation}
    k \ll \omega_1/c_s = 1.56 (2\pi T)\sqrt{3} \approx 17\, T.
\end{equation}
Positing that the UV scale $\Lambdah$ should be at most half of these values, we obtain the estimate
\begin{equation}
    3\, T \lesssim \Lambdah \lesssim 6\, T,
\end{equation}
for realistic to strong coupling.

\subsection{Coupling to quarks and gluons}
\label{sec:QGcoupling}

We will now use the expansion developed in the previous section to couple the fluid to quarks and gluons. This is achieved by furnishing each term of the QCD Lagrangian with a function depending on the dimensionless quantity $X$ \cite{Acanfora:2019con,Matchev:2021fuw}. Additionally, we add all allowed terms containing the current $J^\mu$ together with the fluid Lagrangian $F(X)$, resulting in
\begin{align}
    \mathcal{L} &= F(X) + Z_\psi(X) \Bar{\psi} \frac{i}{2}\gamma^{\mu}\overleftrightarrow{D}_{\mu} \psi- M(X)\Bar{\psi}\psi + Z_A(X) G^a_{\mu\nu} G_a^{\mu\nu} +  \nonumber\\
    &+ P(X) \Bar{\psi} \gamma_{\mu} \psi J^{\mu} + Q(X) \Bar{\psi} \frac{i}{2} \overleftrightarrow{D}_{\mu} \psi J^{\mu}  + R(X) \Bar{\psi} \sigma_{\mu\nu} \psi \omega^{\mu\nu},
    \label{Eq:FullDiracLagrangian}
\end{align}
where we have defined $\omega^{\mu\nu}=\partial^\mu J^\nu - \partial^\nu J^\mu$ and $i \overleftrightarrow{D}=i\overrightarrow{D}-i\overleftarrow{D}$. Note that these are only the relevant and marginal operators of the theory. Irrelevant operators are suppressed by an additional scale $\Lambda_{mfp}>\Lambdah$, related to the breakdown of the fluid picture itself. Thus, when expanding in the fluctuations, the phonon-quark operators and the irrelevant operators generated from the fluid Lagrangian do not mix under renormalization in a mass-independent scheme. One can further reduce the number of operators by using discrete symmetries. For example, $\Bar{\psi} \gamma_{\mu} \psi J^{\mu}$ is not invariant under charge conjugation symmetry and we will therefore exclude it from further considerations.

To determine the coupling of phonons to quarks and gluons we will expand around the equilibrium value as before. In general the terms appearing in the Lagrangian can be classified into three different categories, with each requiring a different treatment in the expansion: The first kind is given by the term only containing the fluid fields, $F(X)$, whose expansion is given by Eqn.~(\ref{Eq:GeneralPhononLagrangian}). The second kind of terms are those of the type $G(X)h(\Bar{\psi},\psi,A_\mu)$ where $G(X)$ is a general function of the scalar $X$ and $h(\Bar{\psi},\psi,A_\mu)$ is function of the other degrees of freedom of the theory. The third and last kind of term has the general form $G(X)h(\Bar{\psi},\psi,A_\mu)v_\mu J^\mu$ where $J^\mu$ is the fluid current and $v_\mu$ is a general vector\footnote{Note that this definition of $v_\mu J^\mu$ also includes $\gamma_\mu J^\mu$.}.

Let us begin with terms of the second type, which we will discuss exemplarily for the term $M(X)\Bar{\psi}\psi$. Similarly to Eqn.~(\ref{Eq:ExpansionF}), the expansion is given by
\begin{align}
    M(X)\Bar{\psi}\psi&= M(1)\Bar{\psi}\psi + 2 b_\mu J_{(1)}^\mu M'(1) \Bar{\psi}\psi \lambda \\
    &+ \left[(J^\mu_{(1)} J_{(1)\mu}+ 2 b_\mu J^\mu_{(2)})M'(1) + 2 (b_\mu J^\mu_{(1)} )^2M''(1) \right] \Bar{\psi}\psi \lambda^2 +\mathcal{O}(\lambda^3).
\end{align}
Note that in contrast to the expansion of $F(X)$, the zeroth-order and linear terms cannot be neglected anymore, since they are multiplied by $\Bar{\psi}\psi$ and therefore are no longer constants or total derivatives. The final expansion of this term is given by
\begin{align}
    M(X)\Bar{\psi}\psi &= \Bar{\psi}\psi\big[M(1) + M^\mu_I \partial_\mu \pi^I + M^{\mu\nu}_{IJ} \partial_\mu \pi^I\partial_\nu \pi^J  + \mathcal{O} ((\partial \pi)^3) \big],
\end{align}
with $M^{\mu\nu}_{IJ}$ having the same form as $F^{\mu\nu}_{IJ}$ after the exchange of $F(X)$ and $M(X)$ and their respective derivatives. Explicitly, the two coefficients are given by 
\begin{align}
    M^\mu_I &= \epsilon^{0\alpha_1 \alpha_2 \mu} \epsilon_{\alpha_1 \alpha_2 I}M'(1) = 2 \delta^{\mu}_{I} M'(1),\\
    M^{\mu\nu}_{IJ} 
    &=\left( 2 \delta^\mu_I \delta^\nu_J - \delta^\mu_0 \delta^\nu_0 \delta_{IJ} - \delta^\nu_I \delta^\mu_J \right) M'(1) + 2 M''(1) \delta^\mu_I \delta^\nu_J .
\end{align}
As before, a natural suppression scale arises after the rescaling Eqn.~(\ref{Eq:PhononRescaling}) of the displacement fields
\begin{align}
    M(X)\Bar{\psi}\psi &= \Bar{\psi}\psi\left[ M(1) + \frac{\Tilde{M}^\mu_I}{\Lambda_h} \partial_\mu \pi^I + \frac{\Tilde{M}^{\mu\nu}_{IJ}}{\Lambda_h^3} \partial_\mu \pi^I\partial_\nu \pi^J + \mathcal{O} \left( \frac{1}{\Lambda_h^5} \right) \right], \label{eq:QuarkPhononInteractionTerm}
\end{align}
with the redefined dimensionless Wilson coefficients
\begin{align}
    \Tilde{M}^\mu_I = \frac{M^\mu_I}{\Lambda_h}  \quad \text{and} \quad    \Tilde{M}^{\mu\nu}_{IJ} = \frac{M^{\mu\nu}_{IJ}}{\Lambda_h}.
\end{align} 
Thus, the operators given in terms of the dimensionless Wilson coefficients are
\begin{align}
    M(X)\bar{\psi}\psi &= M(1)\bar{\psi}\psi + \frac{2 \Tilde{M}'(1)}{\Lambda_h} \bar{\psi}\psi \nabla \cdot \vec{\pi} + \frac{1}{\Lambda_h^3} \left[ 2 \Tilde{M}''(1) (\nabla\cdot\vec{\pi})^2  \right. \nonumber \\
    & \left.+ \left\lbrace (\nabla \cdot \vec{\pi})^2 - 2 (\partial_{t} \vec{\pi})^2 - \partial_I \pi^J \partial_J \pi^I \right\rbrace \Tilde{M}'(1)\right] \bar{\psi} \psi + \mathcal{O}\left( \frac{1}{\Lambda_h^5} \right)
\end{align}

The third kind of terms we have to consider, are the ones where the current $J^\mu$ directly couples to fermionic operators, which we will discuss exemplarily for the term $P(X)\Bar{\psi} \gamma_\mu J^\mu\psi$\footnote{While this term is ruled out by charge conjugation symmetry, we will nevertheless use it for the demonstration of the expansion, since similar terms will be relevant for future works.}.
We again apply the established expansions of $P(X)$ and $J^\mu$, resulting in
\begin{align}
    P(X)\Bar{\psi}\gamma_\mu J^\mu\psi &= \Big[ P(1) + 2 b_\mu J_{(1)}^\mu P'(1) \lambda \nonumber \\
    &+ [(J^\mu_{(1)} J_{(1)\mu}+ 2 b_\mu J^\mu_{(2)})P'(1) + 2 (b_\mu J^\mu_{(1)} )^2P''(1)] \lambda^2 + \mathcal{O}(\lambda^3)\Big] \nonumber \\
    &\times \Bar{\psi} \gamma_\mu \Big[ J^\mu_{(0)} + \lambda J^\mu_{(1)} + \lambda^2 J^\mu_{(2)} + \lambda^3 J^\mu_{(3)} \Big]\psi
\end{align}
Similar to the previous case, fewer simplifications through partial integration arise due to the presence of the extra fields. The final form of the quark-current coupling term, after taking the limit $\lambda \to 1$, then reads as
\begin{align}
    P(X)\Bar{\psi}\gamma_\mu J^\mu \psi &= \Bar{\psi} \gamma_\mu \Big[ P^\mu + P^{\mu\nu}_I \partial_\nu \pi^I + P^{\mu\nu\rho}_{IJ}\partial_\nu \pi^I \partial_\rho \pi^J + \mathcal{O}((\partial\pi)^3)  \Big] \psi,
\end{align}
with the coefficients
\begin{align}
    P^\mu&=b^\mu P(1),\\
    P^{\mu\nu}_I &=\frac{1}{2} \epsilon^{\mu\alpha_1\alpha_2 \nu}\epsilon_{\alpha_1\alpha_2 I} P(1) + b^\mu \epsilon^{0\alpha_1 \alpha_2\nu}\epsilon_{\alpha_1\alpha_2 I} P'(1),\\
    P^{\mu\nu\rho}_{IJ} & = \frac{1}{2} \epsilon^{\mu\alpha_1\nu\rho}\epsilon_{\alpha_1 IJ} P(1) + \frac{1}{2} \epsilon^{\mu \alpha_1 \alpha_2 \nu} \epsilon_{\alpha_1 \alpha_2 I} \epsilon^{0 \alpha_1' \alpha_2' \rho} \epsilon_{\alpha_1' \alpha_2' J} P'(1) \nonumber\\
    &+ b^\mu \left( \frac{1}{4}\epsilon^{0\alpha_1\alpha_2\nu}\epsilon^{0\alpha_1'\alpha_2'\rho}\epsilon_{\alpha_1 \alpha_2 I}\epsilon_{\alpha_1' \alpha_2' J} + \epsilon^{0\alpha_1 \nu\rho} \epsilon_{\alpha_1 IJ} \right)P'(1) \nonumber\\
    &+ \frac{1}{2}b^\mu \epsilon^{0\alpha_1\alpha_2\nu} \epsilon^{0\alpha_1'\alpha_2'\rho} \epsilon_{\alpha_1 \alpha_2 I} \epsilon_{\alpha_1' \alpha_2' J}P''(1).
\end{align}
After introducing the rescaling of the phonon field Eqn.~(\ref{Eq:PhononRescaling}), we recover the suppression through the UV scale
\begin{align}
    P(X)\Bar{\psi}\gamma_\mu J^\mu \psi &= \Bar{\psi} \gamma_\mu \Big[ P^\mu + \frac{\Tilde{P}^{\mu\nu}_I}{\Lambda_h} \partial_\nu \pi^I + \frac{\Tilde{P}^{\mu\nu\rho}_{IJ}}{\Lambda_h^3}\partial_\nu \pi^I \partial_\rho \pi^J + \mathcal{O}((\partial\pi)^3)  \Big] \psi
\end{align}
with
\begin{align}
\Tilde{P}^{\mu\nu}_I = \frac{P^{\mu\nu}_I}{\Lambda_h} \quad \text{and} \quad   \Tilde{P}^{\mu\nu\rho}_{IJ}=\frac{P^{\mu\nu\rho}_{IJ}}{\Lambda_h}.
\end{align}

\begin{figure}[!ht]
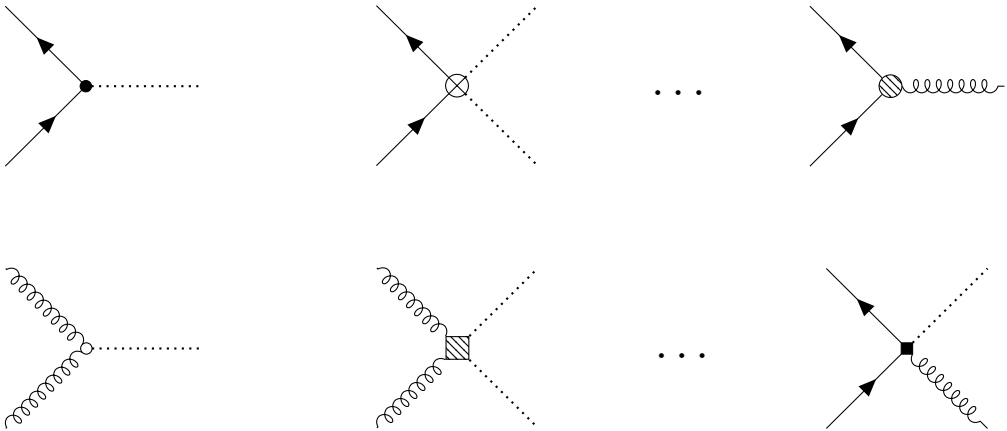

    \begin{subfigure}{0.3\textwidth}
        \centering
        \xusebox{qqPhonon}
    \end{subfigure}
    \begin{subfigure}{0.3\textwidth}
        \centering
        \xusebox{4PointFermionPhonon}
    \end{subfigure}
    \hfill
    $\vcenter{\hbox{\raisebox{4.5em}{\LARGE$\cdots$}}}$
    \hfill
    \begin{subfigure}{0.3\textwidth}
        \centering
        \xusebox{qqbarg}
    \end{subfigure}

    \vspace{1em} 
    
    \begin{subfigure}{0.3\textwidth}
        \centering
        \xusebox{PhononGluon}
    \end{subfigure}
    \begin{subfigure}{0.3\textwidth}
        \centering
        \xusebox{4PointGluonPhonon}
    \end{subfigure}
    \hspace{0.2em}
    $\vcenter{\hbox{\raisebox{4.5em}{\LARGE$\cdots$}}}$
     \hfill
    \begin{subfigure}{0.3\textwidth}
        \centering
        \xusebox{4PointFermionPhononGluon}
    \end{subfigure}
    \caption{Feynman diagrams for quark-phonon, quark-gluon, gluon-phonon, and quark-gluon-phonon interactions. Each diagram exists with an arbitrary number of phonon insertions.}
    \label{fig:FeynmanRules}
\end{figure}

In addition to these phonon-quark interactions, we also have the usual QCD quark-gluon interactions with modified masses and couplings because of the presence of the medium. The here appearing gluons are considered to have momentum larger than the UV scale, since any gluon with smaller momentum is part of the medium and therefore described by the hydrodynamic EFT. Strictly speaking, these gluons are not part of the hierarchy $T < m < \Lambdah$ presented in this work and therefore would allow a formulation of the theory with partial instead of gauge derivatives. The inclusion of these gluons will become important for future works with hierarchy $T < \Lambdah < m$. The hierarchy with the quark mass being the largest scale requires an additional expansion in the mass and would be similar to heavy quark effective theory (HQET). The gluon-phonon and quark-gluon-phonon interactions are suppressed by higher powers of the UV scale compared to the quark-phonon interactions. Some of the possible Feynman diagrams are displayed in Fig.~\ref{fig:FeynmanRules}. Note that due to the expansions performed, an infinite series of diagrams with different numbers of phonon insertions exists for each type of interaction.

\begin{xlrbox}{qqPhonon}
  \begin{tikzpicture}
    \begin{feynman}
      \vertex (a);
      \vertex [dot][right=of a] (b){};
      \vertex [above left=of b] (q1);
      \vertex [below left=of b] (q2);
      \vertex [right=1.5cm of b] (g);

      \diagram*{
        (b) -- [fermion] (q1),
        (b) -- [anti fermion] (q2),
        (b) -- [ghost] (g),
      };
    \end{feynman}
  \end{tikzpicture}
\end{xlrbox}

\begin{xlrbox}{qqbarg}
  \begin{tikzpicture}
    \begin{feynman}
      \vertex (a);
      \vertex [blob, minimum height=0.3cm, minimum width=0.3cm][right=of a] (b){};
      \vertex [above left=of b] (g1);
      \vertex [below left=of b] (g2);
      \vertex [right=1.5cm of b] (g3);

      \diagram*{
        (b) -- [fermion] (g1),
        (b) -- [anti fermion] (g2),
        (b) -- [gluon] (g3),
      };
    \end{feynman}
  \end{tikzpicture}
\end{xlrbox}

\begin{xlrbox}{PhononGluon}
  \begin{tikzpicture}
    \begin{feynman}
      \vertex (a);
      \vertex [empty dot][right=of a] (b){};
      \vertex [above left=of b] (g1);
      \vertex [below left=of b] (g2);
      \vertex [right=1.5cm of b] (g3);

      \diagram*{
        (b) -- [gluon] (g1),
        (b) -- [gluon] (g2),
        (b) -- [ghost] (g3),
      };
    \end{feynman}
  \end{tikzpicture}
\end{xlrbox}

\begin{xlrbox}{4PointFermionPhonon}
  \begin{tikzpicture}
    \begin{feynman}
      \vertex (a);
      \vertex [crossed dot][right=of a] (b){};
      \vertex [above left=of b] (g1);
      \vertex [below left=of b] (g2);
      \vertex [above right=of b] (g3);
      \vertex [below right=of b] (g4);

      \diagram*{
        (b) -- [fermion] (g1),
        (b) -- [anti fermion] (g2),
        (b) -- [ghost] (g3),
        (b) -- [ghost] (g4),
      };
    \end{feynman}
  \end{tikzpicture}
\end{xlrbox}

\begin{xlrbox}{4PointGluonPhonon}
  \begin{tikzpicture}
    \begin{feynman}
      \vertex (a);
      \vertex [blob, minimum height=0.3cm, minimum width=0.3cm, shape=rectangle][right=of a] (b){};
      \vertex [above left=of b] (g1);
      \vertex [below left=of b] (g2);
      \vertex [above right=of b] (g3);
      \vertex [below right=of b] (g4);

      \diagram*{
        (b) -- [gluon] (g1),
        (b) -- [gluon] (g2),
        (b) -- [ghost] (g3),
        (b) -- [ghost] (g4),
      };
    \end{feynman}
  \end{tikzpicture}
\end{xlrbox}

\begin{xlrbox}{4PointFermionPhononGluon}
  \begin{tikzpicture}
    \begin{feynman}
      \vertex (a);
      \vertex [square dot][right=of a] (b){};
      \vertex [above left=of b] (g1);
      \vertex [below left=of b] (g2);
      \vertex [above right=of b] (g3);
      \vertex [below right=of b] (g4);

      \diagram*{
        (b) -- [fermion] (g1),
        (b) -- [anti fermion] (g2),
        (b) -- [ghost] (g3),
        (b) -- [gluon] (g4),
      };
    \end{feynman}
  \end{tikzpicture}
\end{xlrbox}

\section{Quark-phonon scattering}
\label{sec:QPhscattering}

\begin{figure}[!ht]
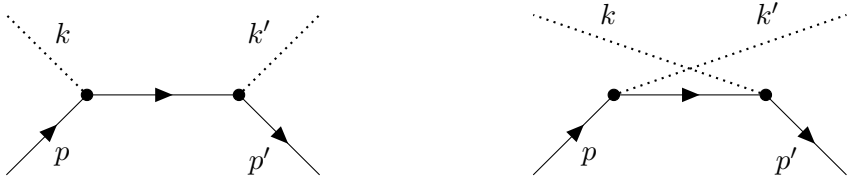

    \begin{subfigure}{0.45\textwidth}
        \centering
        \xusebox{2to2}
    \end{subfigure}
    \begin{subfigure}{0.45\textwidth}
        \centering
        \xusebox{2to2cross}
    \end{subfigure}
    \caption{Feynman diagrams for two-to-two fermion-phonon scattering. The local four-point interaction is suppressed by a higher power of $\Lambdah$ and is therefore neglected.}
    \label{fig:2to2scat}
\end{figure}

In this section, we are going to calculate the $2 \rightarrow 2$ scattering cross-section of the heavy quark with a phonon. The diagrams contributing to this process are given in Fig. \ref{fig:2to2scat}. Using the power counting from Eq.~\ref{eq:QuarkPhononInteractionTerm}, it is clear that these diagrams are suppressed by $\Lambdah^2$. Note that the contribution given by the point interaction between two quarks and two phonons (see middle figure, top row of Fig.~\ref{fig:FeynmanRules}) is suppressed by $\Lambdah^3$, as can also be seen from Eq.~\ref{eq:QuarkPhononInteractionTerm}. The local four point interaction is therefore a subleading term, and we will neglect it in the following.

\begin{xlrbox}{2to2}
  \begin{tikzpicture}
    \begin{feynman}
      \vertex (a);
      \vertex [dot][right=of a] (b){};
      \vertex [above left=of b] (g1);
      \vertex [below left=of b] (q1);
      \vertex [dot][right=2cm of b] (q2){};
      \vertex [above right=of q2] (g2);
      \vertex [below right=of q2] (q3);

      \diagram*{
        (b) -- [ghost, edge label'=\(k\)] (g1),
        (b) -- [anti fermion, edge label=\(p\)] (q1),
        (b) -- [fermion] (q2),
        (q2) -- [ghost, edge label=\(k'\)] (g2),
        (q2) -- [fermion, edge label'=\(p'\)] (q3),
      };
    \end{feynman}
  \end{tikzpicture}
\end{xlrbox}

\begin{xlrbox}{2to2cross}
  \begin{tikzpicture}
    \begin{feynman}
      \vertex (a);
      \vertex [dot][right=of a] (b){};
      \vertex [above left=of b] (g1);
      \vertex [below left=of b] (q1);
      \vertex [dot][right=2cm of b] (q2){};
      \vertex [above right=of q2] (g2);
      \vertex [below right=of q2] (q3);

      \diagram*{
        (q2) -- [ghost, edge label'=\(k\), near end] (g1),
        (b) -- [anti fermion, edge label=\(p\)] (q1),
        (b) -- [fermion] (q2),
        (b) -- [ghost, edge label=\(k'\), near end] (g2),
        (q2) -- [fermion, edge label'=\(p'\)] (q3),
      };
    \end{feynman}
  \end{tikzpicture}
\end{xlrbox}

This scattering process displays interesting phenomenological features, arising from the fact that the phonons are spacelike, i.e. $k^2<0$, and therefore the central heavy quark propagator can go on-shell for certain momentum regions. Thus, one needs to modify the propagator with the width of the unstable particle, which we are going to calculate in the next subsection.

\subsection{Quasiparticle width}
\label{sec:Decay}

\begin{figure}[!ht]
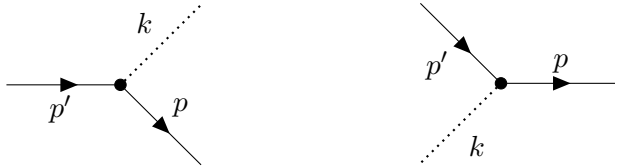

\centering
    \begin{subfigure}{0.35\textwidth}
        \centering
        \xusebox{quarkDecay1}
    \end{subfigure}
    \begin{subfigure}{0.35\textwidth}
        \centering
        \xusebox{quarkDecay2}
    \end{subfigure}
    \caption{Feynman diagrams for two-body quasiparticle width. Left: phonon emission; right: phonon absorption.}
    \label{fig:quarkdecay}
\end{figure}

A quark inside the fluid acquires a finite width for two reasons: A slow-moving quark can absorb a thermal phonon, and a fast-moving quark can emit a phonon. This is possible because phonons have a spacelike dispersion relation. The diagrams for these processes are shown in Fig.~\ref{fig:quarkdecay}.

\begin{xlrbox}{quarkDecay1}
  \begin{tikzpicture}
    \begin{feynman}
      \vertex (a);
      \vertex [dot][left=of a] (b){};
      \vertex [above right=of b] (q1);
      \vertex [below right=of b] (q2);
      \vertex [left=1.5cm of b] (g);

      \diagram*{
        (b) -- [ghost, edge label=\(k\)] (q1),
        (b) -- [fermion, edge label=\(p\)] (q2),
        (b) -- [anti fermion, edge label=\(p'\)] (g),
      };
    \end{feynman}
  \end{tikzpicture}
\end{xlrbox}

\begin{xlrbox}{quarkDecay2}
  \begin{tikzpicture}
    \begin{feynman}
      \vertex (a);
      \vertex [dot][right=of a] (b){};
      \vertex [above left=of b] (q1);
      \vertex [below left=of b] (q2);
      \vertex [right=1.5cm of b] (g);

      \diagram*{
        (b) -- [anti fermion, edge label=\(p'\)] (q1),
        (b) -- [ghost, edge label=\(k\)] (q2),
        (b) -- [fermion, edge label=\(p\)] (g),
      };
    \end{feynman}
  \end{tikzpicture}
\end{xlrbox}

A quark at rest in the fluid can only absorb a phonon; it cannot decay by emission of a phonon. This is different for a quark that moves relative to the fluid at a speed larger than the speed of sound. According to the general formula for two-body decays, the width of the quark with momentum $\vec{p}$ is given by
\begin{align}
    \Gamma(\vec{p}') = & \frac{8\Tilde{M}^{\prime}(1)^2}{\Lambda_h^2} \frac{1}{2E_{p'}} \int \frac{d^3k}{(2\pi)^3 2\omega_k} \frac{d^3p}{(2\pi)^3 2E_p} (2\pi)^4 \delta^{(3)} \left(\vec{p}'-\vec{p}-\vec{k} \right) \delta \left( E_{p'}-E_{p}-\omega_k \right) \nonumber \\
    & \times \left( 2m^2-\omega_k E_{p'}+\vec{p}' \cdot \vec{k} \right) \vec{k}^2 [1+n(\omega_k)] \nonumber \\
    & + \frac{8\Tilde{M}^{\prime}(1)^2}{\Lambda_h^2} \frac{1}{2E_{p'}} \int \frac{d^3k}{(2\pi)^3 2\omega_k} \frac{d^3p}{(2\pi)^3 2E_p} (2\pi)^4 \delta^{(3)} \left(\vec{p}'-\vec{p}+\vec{k} \right) \delta \left( E_{p'}-E_{p}+\omega_k \right) \nonumber \\
    & \times \left( 2m^2+\omega_k E_{p'}-\vec{p}' \cdot \vec{k} \right) \vec{k}^2 \, n(\omega_k),
\end{align}
where $(E',\vec{p}')$ and $(E,\vec{p})$ are the initial and final state four-momenta of the quark, $(\omega,\vec{k}) = (c_sk,\vec{k})$ is the four-momentum of the phonon, and $n(\omega_k)$ is the phonon occupation number. Here $m=M(1)$ represents the in-medium tree-level quark mass. We assume that the quark is on mass-shell before and after the decay and neglect a possible in-medium mass correction.

\begin{figure}[!ht]
\begin{subfigure}[c]{0.48\textwidth}
\centering
\includegraphics[width=1\textwidth]{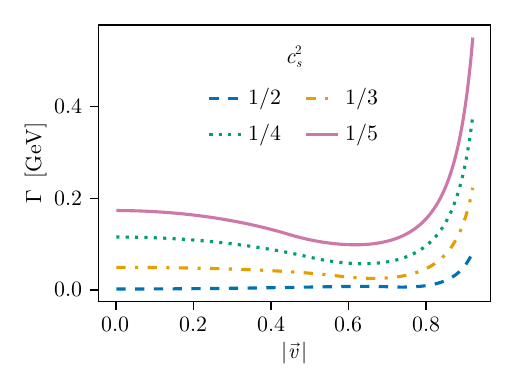}
\subcaption{Quasiparticle width as function of quark velocity for different values of the speed of sound and $T=0.3 \; \mathrm{GeV}$}

\end{subfigure}
\hspace{\fill}
\begin{subfigure}[c]{0.48\textwidth}
\centering
\includegraphics[width=1\textwidth]{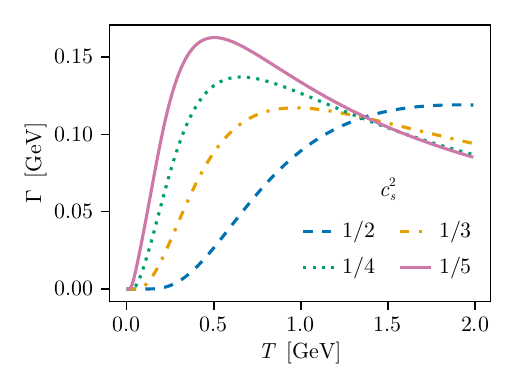}
\subcaption{Quasiparticle width as function of temperature for different values of the speed of sound and $|\vec{p}'|=0.5 \; \mathrm{GeV}$}
\end{subfigure}
\caption{Quasiparticle width as function of heavy quark momentum and temperature for different values of the speed of sound for a charm quark. The quasiparticle width rapidly grows for large quark velocities indicating that the EFT ceases to be valid in this area.}
\label{Fig:DecayWidth}
\end{figure}

After performing the outgoing quark momentum integration and the angular integral for the phonon, we get
\begin{align}
    \Gamma(p') & =  \frac{1}{2\pi E_{p'}} \frac{\Tilde{M}^{\prime}(1)^2}{\Lambda_h^2} \frac{1}{c_s p'}\int_{0}^{\infty} dk\, \left[ 2m^2 k^2 +\frac{1}{2} (1-c_s^2)k^4\right] \theta \left( 2\alpha p'-k\right) [1+n(\omega_k)] \nonumber \\
    & +\frac{1}{2\pi E_{p'}} \frac{\Tilde{M}^{\prime}(1)^2}{\Lambda_h^2} \frac{1}{c_s p'}\int_{0}^{\infty} dk\, \left[ 2m^2 k^2 +\frac{1}{2} (1-c_s^2)k^4\right] \theta \left( 2\beta p'-k\right) \theta \left( 2\alpha p'+k\right) n(\omega_k),
\end{align} \label{eq:Gamma}
where $\alpha=\left( 1-c_s \frac{E_{p'}}{p'} \right)/(1-c_s^2)$ and $\beta=\left( 1+c_s \frac{E_{p'}}{p'} \right)/(1-c_s^2)$. This puts an upper limit on the integral which is non-zero only when $\alpha>0$. Thus the first line, i.e. the emission diagram is non-zero when the Mach condition $v >c_s$ holds, where $v=p'/E_{p'}$, while the emission diagram also has non-zero contributions for smaller velocities. The final radial integral has to be numerically evaluated for the finite temperature piece. The quasiparticle width is plotted in Fig.~\ref{Fig:DecayWidth} for various parameters, with the Wilson coefficient taken as $\tilde{M}^{\prime}(1)=1/2$ here and in the following.

The quasiparticle width displays some interesting features, beginning with a steep rise for large velocities. This behavior is a result of the integration boundaries in Eq.~(\ref{eq:Gamma}) being larger than the UV scale $\Lambdah$. To obtain an estimate for the range of validity of our result we impose that
\begin{align}
    p=mv\gamma \leq \Lambdah &\approx 5T, \\
    \Rightarrow v \leq \frac{\Lambdah}{\sqrt{\Lambdah^2+m^2}} &\approx \frac{5T}{\sqrt{25T^2+m^2}},
\end{align}
resulting in the calculation not being valid for $v \gtrsim 0.8$ for $T=0.3 \;\mathrm{GeV}$ and the charm mass. Owing to its larger mass, the quasiparticle width of the bottom quark diverges already at lower velocities. An increase in the temperature of the medium results in a larger UV scale, leading to a larger range of validity of the quasiparticle width. In the limit of $T \rightarrow 0$ the quasiparticle width is zero if the velocity is smaller than the speed of sound since no resonant state can be formed. For velocities larger than the speed of sound, the quasiparticle width is non-zero since the quark can emit phonons. While initially showing some polynomial behavior, the quasiparticle width decreases as $1/T$ for large temperatures, since the phonon distribution function causes the integrand to scale linearly with the temperature, and the quasiparticle width is proportional to $\Lambdah^{-2} \propto T^{-2}$ resulting in an overall $1/T$ scaling.

\begin{figure}
    \centering
    \includegraphics[width=0.8\linewidth]{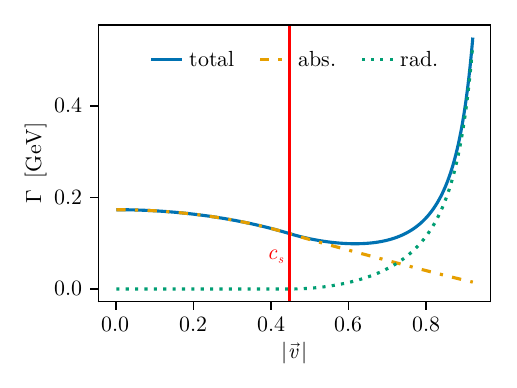}
    \caption{Radiation and absorption contributions to the quasiparticle width as function of the heavy quark velocity for $T=0.3 \; \mathrm{GeV}$ and $c_s^2=1/5$. The radiation contribution is kinematically only allowed if the heavy quark is faster than the photons of the medium. For large velocities the quasiparticle width rises steeply, because the emitted phonons have a momentum larger than the UV scale of the theory, resulting in the EFT not being valid anymore.}
    \label{fig:Gamma contributions}
\end{figure}

A similar picture arises when considering the quasiparticle width as function of the heavy quark velocity, where the emission of a phonon is only possible if the heavy quark is faster than the speed of sound, leading to a larger quasiparticle width for higher velocities. For a larger velocity of the heavy quark, the emitted phonon also becomes more energetic, resulting in the steep rise of the quasiparticle width, since the emitted phonons with momentum larger than the UV scale are formally not part of the theory and therefore lead to unitarity violations (yellow line, Fig.~\ref{fig:Gamma contributions}). For velocities of the heavy quark below the speed of sound only absorption of a thermal phonon is possible (green line, Fig.~\ref{fig:Gamma contributions}), which becomes less efficient at large velocities. This interplay between the absorption and emission mechanisms leads to the rise and fall of the quasiparticle width as function of the velocity. Larger values of the speed of sound hereby result in the phonon being closer to a photon or a gluon in its dispersion, effectively leading to a smaller quasiparticle width.

\subsection{Scattering matrix elements}
\label{sec:Scattering}

In the case of the quark-phonon scattering the matrix elements corresponding to the two diagrams shown in Figure \ref{fig:2to2scat} are 
\begin{eqnarray}
    i \mathcal{A}_{1} &= i\frac{4\Tilde{M}^{\prime}(1)^2}{\Lambda_h^2} \bar{u}(p^\prime)(\slashed{p}+\slashed{k}+m)u(p)\frac{k_{i}\varepsilon^{i}(k) k^{\prime}_{j}\varepsilon^{*j}(k^\prime)}{2p \cdot k + k^2 +im \Gamma(p+k)}, \\
    i \mathcal{A}_{2} &= i\frac{4\Tilde{M}^{\prime}(1)^2}{\Lambda_h^2} \bar{u}(p^\prime)(\slashed{p}-\slashed{k}^{\prime}+m)u(p)\frac{k_{i}\varepsilon^{i}(k) k^{\prime}_{j}\varepsilon^{*j}(k^\prime)}{-2p \cdot k^{\prime} + k^{\prime 2}+im \Gamma(p-k')}.
\end{eqnarray}

After carrying out the spin and polarization sums, the scattering matrix element is given by
\begin{align}
     |\mathcal{A}|^2 =& |\mathcal{A}_{1}+\mathcal{A}_2|^2 \nonumber \\
     =& \frac{16\Tilde{M}^{'}(1)^4}{\Lambda_h^4} \left\lbrace 4 \frac{4 m^2 \left( k \cdot p'+ p \cdot p' +m^2 \right)+2 \left(k \cdot p\right) \left( k \cdot p' +2 m^2\right)+k^2 \left(m^2-p \cdot p' \right)}{|2p \cdot k + k^2+i m\Gamma(p+k)|^2}  \right. \nonumber \\
     & \left. +4 \frac{4 m^2 \left( -k' \cdot p' +p \cdot p' +m^2 \right) +2 \left(k' \cdot p \right) \left( k' \cdot p' -2 m^2\right)+k'^ {2} \left(m^2-p \cdot p' \right)}{|-2p \cdot k' + k'^2+i m\Gamma(p-k')|^2}  \right. \nonumber \\
     & \left.  + 4  \frac{(k \cdot p) ( 2m^2 - k' \cdot p')+(k \cdot k') (p \cdot p'-m^2) -(k \cdot p')(k' \cdot p)}{[2p \cdot k + k^2 +im\Gamma(p+k)][-2 p\cdot k' + k'^2+im\Gamma(p-k')]^{*}} \right. \nonumber \\
     & \left. +4\frac{2m^2 (k \cdot p'-k' \cdot p-k' \cdot p'+2 p \cdot p')+4m^4 }{[2p \cdot k + k^2 +im\Gamma(p+k)][-2 p\cdot k' + k'^2+im\Gamma(p-k')]^{*}} \right. \nonumber \\
     & \left.  + 4  \frac{(k \cdot p) ( 2m^2 - k' \cdot p')+(k \cdot k') (p \cdot p'-m^2) -(k \cdot p')(k' \cdot p)}{[2p \cdot k + k^2 +im\Gamma(p+k)]^{*}[-2 p\cdot k' + k'^2+im\Gamma(p-k')]} \right. \nonumber \\
     & \left. +4\frac{2m^2 (k \cdot p'-k' \cdot p-k' \cdot p'+2 p \cdot p')+4m^4 }{[2p \cdot k + k^2 +im\Gamma(p+k)]^{*}[-2 p\cdot k' + k'^2+im\Gamma(p-k')]}\right\rbrace \vec{k}^2 \vec{k'}^2 \label{eq:SquaredMatrixElement}
\end{align}

With this, the cross section of the process Fig.~\ref{fig:2to2scat} is given by \footnote{Note that the velocity vector $\vec{v}_k$ for a phonon is not given by $\vec{k}/E_k$, but rather by $\vec{v_k}=c_s \vec{k}/|\vec{k}|$. }

\begin{align}
    \sigma = \frac{1}{2E_p 2E_k} \frac{1}{|\vec{v}_p-\vec{v}_k|} \int \frac{\mathrm{d}^3 k'}{(2\pi)^3 2 E_{k'}} \frac{\mathrm{d}^3 p'}{(2\pi)^3 2 E_{p'}} |\mathcal{A}|^2 (2\pi)^4 \delta^{(4)}(p+k-p'-k') \left[1 + n_B(k') \right].
\end{align}
We evaluate the four delta function enforcing energy-momentum conservation by employing the relation
\begin{align}
    \int \frac{\mathrm{d}^3 p'}{2E_p'} = \int \mathrm{d}^4 p' \delta (p'^2-m^2) \theta(E_{p'}),
\end{align}
and carrying out the integral over $p'$. 
The resulting delta function is then eliminated by the transformation to spherical coordinates $(k'_x,k'_y,k'_z) \to |\vec{k}'|(\sin \theta_{k'} \cos \phi_{k'}, \sin \theta_{k'} \sin \phi_{k'}, \cos \theta_{k'})$ and the integration over $|\vec{k}'|$. The remaining integrals over the angles $\phi_{k'}$ and $\theta_{k'}$ are carried out using an adaptive Gauss-Kronrod quadrature.

\begin{figure}[!ht]
\begin{subfigure}[c]{0.5\textwidth}
\includegraphics[width=1\textwidth]{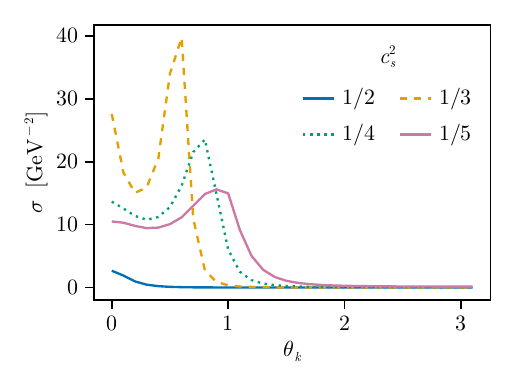}
\subcaption{Scattering cross section as function of the phonon polar angle for different speeds of sound values for a charm quark.}
\end{subfigure}
\hspace{0.5cm}
\begin{subfigure}[c]{0.5\textwidth}
\includegraphics[width=1\textwidth]{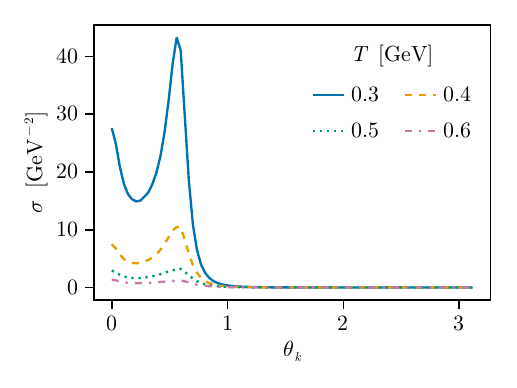}
\subcaption{Scattering cross section as function of the phonon polar angle for different temperatures for charm quark.\phantom{temperature temperature}}
\end{subfigure}
\caption{Scattering cross section as function of the phonon polar angle for different values of the speed of sound (left) and temperature (right) for a charm quark. The momentum of the charm quark is chosen to be along the $z$-direction with magnitude $|\vec{p}|=1 \; \mathrm{GeV}$ ($|\vec{v}|\approx 0.62$). The fixed temperature, phonon momentum and speed of sound for the left and right panels are $T=0.3 \; \mathrm{GeV}$, $|\vec{k}|=0.3 \; \mathrm{GeV}$ and $c_s^2=1/3$, respectively. The cross section scales inversely with the temperature, since $\Lambdah \propto T$. The broadening of the peak in the scattering cross section for smaller speeds of sound can be attributed to the larger quasiparticle width.}
\label{Fig:CrossSection}
\end{figure}

\begin{figure}[h]
    \centering
    \includegraphics[width=0.8\linewidth]{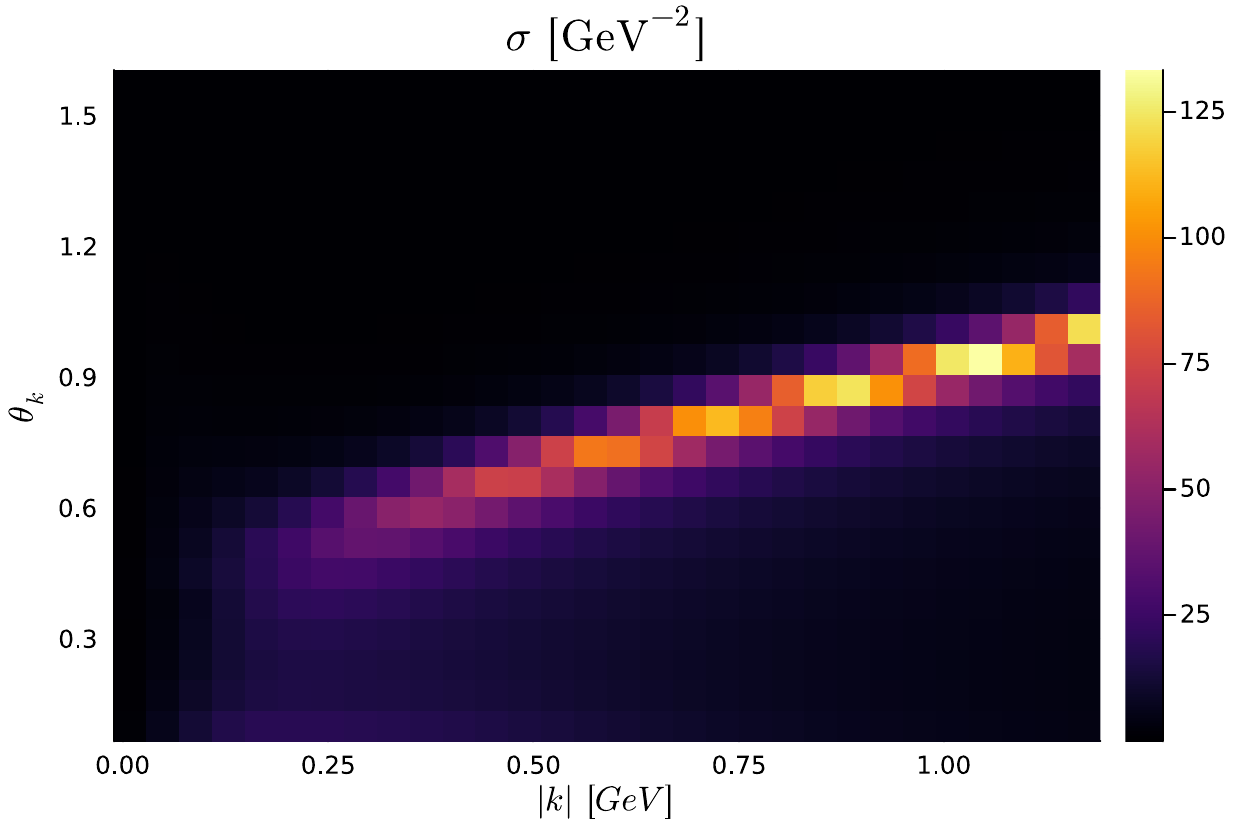}
    \caption{Scattering cross section as function of the phonon momentum and polar angle for a charm quark aligned with the $z$-axis and momentum $|\vec{p}|=1 \; \mathrm{GeV}$. The temperature and speed of sound of the medium are given by $T =0.3 \; \mathrm{GeV}$ and $c_s^2 = 1/3$, respectively. For larger phonon momenta, the ridge at which the intermediate state in the scattering goes on-shell and forms an excited state can be seen.}
    \label{fig:Kinematics_Sigma}
\end{figure}

The resulting scattering cross section as function of the phonon polar angle for different values of the speed of sound and medium temperature is displayed in Fig.~\ref{Fig:CrossSection}. The heavy quark is here taken to be a charm quark, with its momentum being aligned along the $z$-axis and having the magnitude $|\vec{p}| = 1 \; \mathrm{GeV}$ ($|\vec{v}|\approx 0.62$). The fixed temperature, phonon momentum and speed of sound for the left and right panels are $T=0.3 \; \mathrm{GeV}$, $|\vec{k}|=0.3 \; \mathrm{GeV}$ and $c_s^2=1/3$, respectively. The Wilson coefficient is taken to be $\tilde{M}^{\prime}(1)=1/2$ in Eq.~(\ref{eq:SquaredMatrixElement}) and Eq.~(\ref{eq:Gamma}). The scattering cross section hereby displays interesting features:

Since the heavy quark is aligned with the $z$-axis, the polar angle of the phonon effectively serves as scattering angle between the two. The cross section vanishes for large angles (i.e. a head-on collision), but is non-vanishing and rather enhanced for the collinear limit of the heavy quark and phonon ($\theta_k=0$). This behavior is explained by considering the interplay between interaction time and the energy of the phonons. For a head-on collision the heavy quark and phonon interact on a much shorter time scale, compared to the collinear limit. Therefore, a more energetic phonon is required to change the quark's momentum, which is suppressed by the thermal distribution and the UV scale. In the collinear limit the interaction time is much larger, resulting in a non-vanishing cross section. In particular, the cross-section gets enhanced for heavy quarks moving with a speed close to the speed of sound, and even diverging for the exact same speed, in the collinear limit, reminiscent of the usual collinear divergence. In between, the intermediate particle of the scattering process goes on-shell for a speed of sound and phonon momentum-dependent angle, leading to a resonance peak. The height of this peak scales inversely with the temperature, since the integral of the scattering cross section results in a polynomial of leading term $\propto T^3$, resulting in an overall $1/T$ dependence through the inclusion of the UV scale $1/\Lambdah^4 \propto 1/T^4$. For different speeds of sound the cross section changes more dramatically, since the position of the resonance and the quasiparticle width depend on the speed of sound. This results in a wider peak at larger angles for smaller speeds of sound, since the quasiparticle width is also larger for smaller values of $c_s$. Specifically, the resonance appears at smaller angles for larger values of the speed of sound at fixed phonon momentum. Therefore, for $c_s^2=1/2$ (blue curve, left panel Fig.~\ref{Fig:CrossSection}) the resonance condition cannot be met, resulting in the strongly suppressed interaction cross section for $|\vec{k}| = 0.3 \; \mathrm{GeV}$. As seen below, this resonance peak can be recovered at larger phonon momenta (while Fig.~\ref{fig:Kinematics_Sigma} is for $c_s^2=1/3$ instead of $c_s^2=1/2$, the general characteristics of the curves are the same).


We also plot the scattering cross-section in the $(|k|,\theta_k)$ plane in Fig.~\ref{fig:Kinematics_Sigma}. Just like in Fig. \ref{Fig:CrossSection}, the peak corresponds to angles where the intermediate state goes on-shell with the additional effect that this angle changes as a function of the incoming phonon momenta. Thus, this gives rise to the band like structure. With larger $k$, the peak is sharper since quasiparticle width goes down.

\section{Conclusion}
\label{sec:Conclusions}

In this work, we presented first steps towards the description of the interactions between the strongly coupled quark gluon plasma and a heavy quark in terms of an effective field theory. This was achieved by coupling the heavy quark to three scalar fields which represent the fluid degrees of freedom of the QGP. We then derived the interactions between the collective longitudinal excitations of these scalar fields (phonons) and the heavy quark as a systematic expansion in the UV scale with appropriate Wilson coefficients. Within this framework, the effective degrees of freedom are weakly interacting and thus make the formalism amenable to perturbative methods.

In the second half of this paper, we used the fluid EFT framework to determine the two-to-two scattering cross section between a heavy quark and a phonon. Quantitative predictions will require the determination of the relevant Wilson coefficients. These can be obtained either by fitting observables to experimental data or by a matching calculation based on the microscopic theory. We plan to determine the Wilson coefficient $\tilde{M}'(1)$ by matching it to strongly coupled AdS/CFT results for observables that can be calculated in both frameworks. Similarly, the Wilson coefficients for the gluon-phonon coupling can be determined by matching to momentum broadening calculations. Additional constraints for the quark-phonon-gluon coupling can arise from jet observables. It is also conceivable to obtain stronger constraints of the structure of the Lagrangian itself through matching to other top-down EFTs (like number conservation and charge conjugation for heavy quarks in HQET). With the necessary Wilson coefficients determined, QHEFT has predictive power and will be used to calculate quantities required for the comparison with experimental data, such as the Boltzmann collision kernel of the heavy quark distribution or the heavy quark drag coefficients appearing in a Fokker-Planck equation.

\acknowledgments

We would like to thank Marston Copeland, Noah Koliadko and Ira Rothstein for useful discussions. We would also like to thank Yu Fu for feedback on the manuscript. This work was supported by the grant DE-FG02-05ER41367 from the U.S. Department of Energy, Office of Science, Nuclear Physics. JR was supported by the U.S. Department of Energy through the Topical Collaboration in Nuclear Theory on Heavy-Flavor Theory (HEFTY) for QCD Matter under award no. DE-SC0023547. CS is supported in part by the National Science Foundation under grant numbers ACI-1550225 and OAC-2004571 (CSSI:X-SCAPE) within the framework of the JETSCAPE collaboration.

\bibliographystyle{JHEP}
\bibliography{biblio.bib}

@article{Endlich:2010hf,
    author = "Endlich, Solomon and Nicolis, Alberto and Rattazzi, Riccardo and Wang, Junpu",
    title = "{The Quantum mechanics of perfect fluids}",
    eprint = "1011.6396",
    archivePrefix = "arXiv",
    primaryClass = "hep-th",
    doi = "10.1007/JHEP04(2011)102",
    journal = "JHEP",
    volume = "04",
    pages = "102",
    year = "2011"
}

@article{Dubovsky:2011sj,
    author = "Dubovsky, Sergei and Hui, Lam and Nicolis, Alberto and Son, Dam Thanh",
    title = "{Effective field theory for hydrodynamics: thermodynamics, and the derivative expansion}",
    eprint = "1107.0731",
    archivePrefix = "arXiv",
    primaryClass = "hep-th",
    doi = "10.1103/PhysRevD.85.085029",
    journal = "Phys. Rev. D",
    volume = "85",
    pages = "085029",
    year = "2012"
}

@article{Grozdanov:2013dba,
    author = "Grozdanov, Sa{\v{s}}o and Polonyi, Janos",
    title = "{Viscosity and dissipative hydrodynamics from effective field theory}",
    eprint = "1305.3670",
    archivePrefix = "arXiv",
    primaryClass = "hep-th",
    reportNumber = "OUTP-13-10P",
    doi = "10.1103/PhysRevD.91.105031",
    journal = "Phys. Rev. D",
    volume = "91",
    number = "10",
    pages = "105031",
    year = "2015"
}

@article{Buchel:2008sh,
    author = "Buchel, Alex",
    title = "{Resolving disagreement for eta/s in a CFT plasma at finite coupling}",
    eprint = "0805.2683",
    archivePrefix = "arXiv",
    primaryClass = "hep-th",
    reportNumber = "UWO-TH-08-8",
    doi = "10.1016/j.nuclphysb.2008.05.024",
    journal = "Nucl. Phys. B",
    volume = "803",
    pages = "166--170",
    year = "2008"
}

@article{Liu:2018kfw,
    author = "Liu, Hong and Glorioso, Paolo",
    title = "{Lectures on non-equilibrium effective field theories and fluctuating hydrodynamics}",
    eprint = "1805.09331",
    archivePrefix = "arXiv",
    primaryClass = "hep-th",
    reportNumber = "MIT-CTP/5018; EFI-18-8, MIT-CTP-5018, EFI-18-8",
    doi = "10.22323/1.305.0008",
    journal = "PoS",
    volume = "TASI2017",
    pages = "008",
    year = "2018"
}

@article{Haehl:2018lcu,
    author = "Haehl, Felix M. and Loganayagam, R. and Rangamani, Mukund",
    title = "{Effective Action for Relativistic Hydrodynamics: Fluctuations, Dissipation, and Entropy Inflow}",
    eprint = "1803.11155",
    archivePrefix = "arXiv",
    primaryClass = "hep-th",
    doi = "10.1007/JHEP10(2018)194",
    journal = "JHEP",
    volume = "10",
    pages = "194",
    year = "2018"
}

@article{Montenegro:2016gjq,
    author = "Montenegro, David and Torrieri, Giorgio",
    title = "{Lagrangian formulation of relativistic Israel-Stewart hydrodynamics}",
    eprint = "1604.05291",
    archivePrefix = "arXiv",
    primaryClass = "hep-th",
    doi = "10.1103/PhysRevD.94.065042",
    journal = "Phys. Rev. D",
    volume = "94",
    number = "6",
    pages = "065042",
    year = "2016"
}

@article{Gupta:2023tue,
    author = "Gupta, Sourendu",
    title = "{Spin polarization of heavy quarks in matter: predictions from effective field theories}",
    eprint = "2307.12250",
    archivePrefix = "arXiv",
    primaryClass = "hep-ph",
    month = "7",
    year = "2023"
}

@article{Stoecker:2004qu,
    author = "Stoecker, Horst",
    editor = "Rischke, D. and Levin, G.",
    title = "{Collective flow signals the quark gluon plasma}",
    eprint = "nucl-th/0406018",
    archivePrefix = "arXiv",
    doi = "10.1016/j.nuclphysa.2004.12.074",
    journal = "Nucl. Phys. A",
    volume = "750",
    pages = "121--147",
    year = "2005"
}

@article{Casalderrey-Solana:2004fdk,
    author = "Casalderrey-Solana, J. and Shuryak, E. V. and Teaney, D.",
    editor = "Roland, Gunther and Trainor, Tom",
    title = "{Conical flow induced by quenched QCD jets}",
    eprint = "hep-ph/0411315",
    archivePrefix = "arXiv",
    doi = "10.1088/1742-6596/27/1/003",
    journal = "J. Phys. Conf. Ser.",
    volume = "27",
    pages = "22--31",
    year = "2005"
}

@article{Ruppert:2005uz,
    author = "Ruppert, Jorg and Muller, Berndt",
    title = "{Waking the colored plasma}",
    eprint = "hep-ph/0503158",
    archivePrefix = "arXiv",
    doi = "10.1016/j.physletb.2005.04.075",
    journal = "Phys. Lett. B",
    volume = "618",
    pages = "123--130",
    year = "2005"
}

@article{Renk:2005si,
    author = "Renk, Thorsten and Ruppert, Jorg",
    title = "{Mach cones in an evolving medium}",
    eprint = "hep-ph/0509036",
    archivePrefix = "arXiv",
    reportNumber = "DUKE-TH-05-279",
    doi = "10.1103/PhysRevC.73.011901",
    journal = "Phys. Rev. C",
    volume = "73",
    pages = "011901",
    year = "2006"
}

@article{Friess:2006fk,
    author = "Friess, Joshua J. and Gubser, Steven S. and Michalogiorgakis, Georgios and Pufu, Silviu S.",
    title = "{The Stress tensor of a quark moving through N=4 thermal plasma}",
    eprint = "hep-th/0607022",
    archivePrefix = "arXiv",
    reportNumber = "PUPT-2201",
    doi = "10.1103/PhysRevD.75.106003",
    journal = "Phys. Rev. D",
    volume = "75",
    pages = "106003",
    year = "2007"
}

@article{Gubser:2007ga,
    author = "Gubser, Steven S. and Pufu, Silviu S. and Yarom, Amos",
    title = "{Sonic booms and diffusion wakes generated by a heavy quark in thermal AdS/CFT}",
    eprint = "0706.4307",
    archivePrefix = "arXiv",
    primaryClass = "hep-th",
    reportNumber = "PUPT-2239, LMU-ASC-42-07",
    doi = "10.1103/PhysRevLett.100.012301",
    journal = "Phys. Rev. Lett.",
    volume = "100",
    pages = "012301",
    year = "2008"
}

@article{Chesler:2007sv,
    author = "Chesler, Paul M. and Yaffe, Laurence G.",
    title = "{The Stress-energy tensor of a quark moving through a strongly-coupled N=4 supersymmetric Yang-Mills plasma: Comparing hydrodynamics and AdS/CFT}",
    eprint = "0712.0050",
    archivePrefix = "arXiv",
    primaryClass = "hep-th",
    doi = "10.1103/PhysRevD.78.045013",
    journal = "Phys. Rev. D",
    volume = "78",
    pages = "045013",
    year = "2008"
}

@article{Neufeld:2008hs,
    author = "Neufeld, R. B.",
    title = "{Fast Partons as a Source of Energy and Momentum in a Perturbative Quark-Gluon Plasma}",
    eprint = "0805.0385",
    archivePrefix = "arXiv",
    primaryClass = "hep-ph",
    doi = "10.1103/PhysRevD.78.085015",
    journal = "Phys. Rev. D",
    volume = "78",
    pages = "085015",
    year = "2008"
}

@article{Neufeld:2010xi,
    author = "Neufeld, R. B.",
    title = "{Thermal field theory derivation of the source term induced by a fast parton from the quark energy-momentum tensor}",
    eprint = "1011.4979",
    archivePrefix = "arXiv",
    primaryClass = "hep-ph",
    doi = "10.1103/PhysRevD.83.065012",
    journal = "Phys. Rev. D",
    volume = "83",
    pages = "065012",
    year = "2011"
}

@article{Chaudhuri:2005vc,
    author = "Chaudhuri, A. K. and Heinz, Ulrich",
    title = "{Effect of jet quenching on the hydrodynamical evolution of QGP}",
    eprint = "nucl-th/0503028",
    archivePrefix = "arXiv",
    doi = "10.1103/PhysRevLett.97.062301",
    journal = "Phys. Rev. Lett.",
    volume = "97",
    pages = "062301",
    year = "2006"
}

@article{Betz:2010qh,
    author = "Betz, Barbara and Noronha, Jorge and Torrieri, Giorgio and Gyulassy, Miklos and Rischke, Dirk H.",
    title = "{Universal Flow-Driven Conical Emission in Ultrarelativistic Heavy-Ion Collisions}",
    eprint = "1005.5461",
    archivePrefix = "arXiv",
    primaryClass = "nucl-th",
    doi = "10.1103/PhysRevLett.105.222301",
    journal = "Phys. Rev. Lett.",
    volume = "105",
    pages = "222301",
    year = "2010"
}

@article{Bouras:2014rea,
    author = "Bouras, I. and Betz, B. and Xu, Z. and Greiner, C.",
    title = "{Mach cones in viscous heavy-ion collisions}",
    eprint = "1401.3019",
    archivePrefix = "arXiv",
    primaryClass = "hep-ph",
    doi = "10.1103/PhysRevC.90.024904",
    journal = "Phys. Rev. C",
    volume = "90",
    number = "2",
    pages = "024904",
    year = "2014"
}

@article{Casalderrey-Solana:2020rsj,
    author = "Casalderrey-Solana, Jorge and Milhano, Jos{\'e} Guilherme and Pablos, Daniel and Rajagopal, Krishna and Yao, Xiaojun",
    title = "{Jet Wake from Linearized Hydrodynamics}",
    eprint = "2010.01140",
    archivePrefix = "arXiv",
    primaryClass = "hep-ph",
    reportNumber = "MIT-CTP/5246",
    doi = "10.1007/JHEP05(2021)230",
    journal = "JHEP",
    volume = "05",
    pages = "230",
    year = "2021"
}

@article{Yang:2021qtl,
    author = "Yang, Zhong and Chen, Wei and He, Yayun and Ke, Weiyao and Pang, Longgang and Wang, Xin-Nian",
    title = "{Search for the Elusive Jet-Induced Diffusion Wake in $Z/\gamma$-Jets with 2D Jet Tomography in High-Energy Heavy-Ion Collisions}",
    eprint = "2101.05422",
    archivePrefix = "arXiv",
    primaryClass = "hep-ph",
    doi = "10.1103/PhysRevLett.127.082301",
    journal = "Phys. Rev. Lett.",
    volume = "127",
    number = "8",
    pages = "082301",
    year = "2021"
}

@article{CMS:2025dua,
    author = "Chekhovsky, Vladimir and others",
    collaboration = "CMS",
    title = "{Evidence of medium response to hard probes using correlations of Z bosons with hadrons in heavy ion collisions}",
    eprint = "2507.09307",
    archivePrefix = "arXiv",
    primaryClass = "nucl-ex",
    reportNumber = "CMS-HIN-23-006, CERN-EP-2025-046",
    month = "7",
    year = "2025"
}

@article{Ghiglieri:2018dib,
    author = "Ghiglieri, Jacopo and Moore, Guy D. and Teaney, Derek",
    title = "{QCD Shear Viscosity at (almost) NLO}",
    eprint = "1802.09535",
    archivePrefix = "arXiv",
    primaryClass = "hep-ph",
    reportNumber = "CERN-TH-2018-009",
    doi = "10.1007/JHEP03(2018)179",
    journal = "JHEP",
    volume = "03",
    pages = "179",
    year = "2018"
}

@article{Andersen:2014dua,
    author = "Andersen, Jens O. and Haque, Najmul and Mustafa, Munshi G. and Strickland, Michael and Su, Nan",
    editor = "Andrianov, Alexander and Brambilla, Nora and Kim, Victor and Kolevatov, Sergei",
    title = "{Equation of State for QCD at finite temperature and density. Resummation versus lattice data}",
    eprint = "1411.1253",
    archivePrefix = "arXiv",
    primaryClass = "hep-ph",
    doi = "10.1063/1.4938592",
    journal = "AIP Conf. Proc.",
    volume = "1701",
    number = "1",
    pages = "020003",
    year = "2016"
}

@article{Waeber:2015oka,
    author = {Waeber, Sebastian and Sch{\"a}fer, Andreas and Vuorinen, Aleksi and Yaffe, Laurence G.},
    title = "{Finite coupling corrections to holographic predictions for hot QCD}",
    eprint = "1509.02983",
    archivePrefix = "arXiv",
    primaryClass = "hep-th",
    reportNumber = "HIP-2015-26-TH, INT-PUB-15-047",
    doi = "10.1007/JHEP11(2015)087",
    journal = "JHEP",
    volume = "11",
    pages = "087",
    year = "2015"
}

@article{Arnold:2002zm,
    author = "Arnold, Peter Brockway and Moore, Guy D. and Yaffe, Laurence G.",
    title = "{Effective kinetic theory for high temperature gauge theories}",
    eprint = "hep-ph/0209353",
    archivePrefix = "arXiv",
    doi = "10.1088/1126-6708/2003/01/030",
    journal = "JHEP",
    volume = "01",
    pages = "030",
    year = "2003"
}

@article{Kovtun:2005ev,
    author = "Kovtun, Pavel K. and Starinets, Andrei O.",
    title = "{Quasinormal modes and holography}",
    eprint = "hep-th/0506184",
    archivePrefix = "arXiv",
    reportNumber = "NSF-KITP-05-41",
    doi = "10.1103/PhysRevD.72.086009",
    journal = "Phys. Rev. D",
    volume = "72",
    pages = "086009",
    year = "2005"
}

@article{Berti:2009kk,
    author = "Berti, Emanuele and Cardoso, Vitor and Starinets, Andrei O.",
    title = "{Quasinormal modes of black holes and black branes}",
    eprint = "0905.2975",
    archivePrefix = "arXiv",
    primaryClass = "gr-qc",
    doi = "10.1088/0264-9381/26/16/163001",
    journal = "Class. Quant. Grav.",
    volume = "26",
    pages = "163001",
    year = "2009"
}

@article{DeLescluze:2025jqx,
    author = "De Lescluze, Matisse and Heller, Michal P.",
    title = "{Quasinormal Modes of Nonthermal Fixed Points}",
    eprint = "2502.01622",
    archivePrefix = "arXiv",
    primaryClass = "hep-th",
    doi = "10.1103/tz78-jfbw",
    journal = "Phys. Rev. Lett.",
    volume = "135",
    number = "9",
    pages = "091601",
    year = "2025"
}

@article{Acanfora:2019con,
    author = "Acanfora, Francesca and Esposito, Angelo and Polosa, Antonio D.",
    title = "{Sub-GeV Dark Matter in Superfluid He-4: an Effective Theory Approach}",
    eprint = "1902.02361",
    archivePrefix = "arXiv",
    primaryClass = "hep-ph",
    doi = "10.1140/epjc/s10052-019-7057-0",
    journal = "Eur. Phys. J. C",
    volume = "79",
    number = "7",
    pages = "549",
    year = "2019"
}

@article{Matchev:2021fuw,
    author = "Matchev, Konstantin T. and Smolinsky, Jordan and Xue, Wei and You, Yining",
    title = "{Superfluid effective field theory for dark matter direct detection}",
    eprint = "2108.07275",
    archivePrefix = "arXiv",
    primaryClass = "hep-ph",
    doi = "10.1007/JHEP05(2022)034",
    journal = "JHEP",
    volume = "05",
    pages = "034",
    year = "2022"
}

@article{Yang:2022nei,
    author = "Yang, Zhong and Luo, Tan and Chen, Wei and Pang, Long-Gang and Wang, Xin-Nian",
    title = "{3D Structure of Jet-Induced Diffusion Wake in an Expanding Quark-Gluon Plasma}",
    eprint = "2203.03683",
    archivePrefix = "arXiv",
    primaryClass = "hep-ph",
    doi = "10.1103/PhysRevLett.130.052301",
    journal = "Phys. Rev. Lett.",
    volume = "130",
    number = "5",
    pages = "052301",
    year = "2023"
}
\end{document}